\documentclass[usenatbib]{mnras}
\pdfoutput=1
\usepackage{amssymb}
\usepackage{amsmath}
\usepackage{natbib}
\usepackage[pdftex]{graphicx}
\usepackage{subfigure}
\usepackage{booktabs}
\usepackage{tabularx}
\usepackage{float}
\usepackage{afterpage}
\usepackage{pdflscape}

\begin{document}
\label{firstpage}
\title[Tides and convection]
{On the interaction between fast tides and convection}
 \author[A. J. Barker \& A. A. V. Astoul]{
 Adrian. J. Barker\thanks{Email address: A.J.Barker@leeds.ac.uk} and 
 Aur\'{e}lie A. V. Astoul
 \\ Department of Applied Mathematics, School of Mathematics, University of Leeds, Leeds, LS2 9JT, UK}

\pagerange{\pageref{firstpage}--\pageref{lastpage}}
\maketitle
\pubyear{2021}

\begin{abstract}
The interaction between equilibrium tides and convection in stellar envelopes is often considered important for tidal evolution in close binary and extrasolar planetary systems. Its efficiency for fast tides has however long been controversial, when the tidal frequency exceeds the turnover frequency of convective eddies. Recent numerical simulations indicate that convection can act like an effective viscosity which decays quadratically with tidal frequency for fast tides, resulting in inefficient dissipation in many applications involving pre- and main-sequence stars and giant planets. A new idea was however recently proposed by Terquem (2021), who suggested Reynolds stresses involving correlations between tidal flow components dominate the interaction instead of correlations between convective flow components as usually assumed. They further showed that this can potentially significantly enhance tidal dissipation for fast tides in many applications. Motivated by the importance of this problem for tidal dissipation in stars and planets, we directly compute this new term using analytical arguments and global spherical simulations using Boussinesq and anelastic hydrodynamic models. We demonstrate that the new term proposed by Terquem vanishes identically for equilibrium tides interacting with convection in both Boussinesq and anelastic models; it is therefore unlikely to contribute to tidal dissipation in stars and planets.
\end{abstract}


\section{Introduction}
\label{Intro}

The action of turbulent convection in damping equilibrium tides in stellar envelopes has long been thought to be one of the dominant mechanisms of tidal dissipation, and has been proposed to circularize binary orbits and synchronize their spins, as well as contribute to short-period planetary orbital evolution \citep[e.g.][]{Zahn1989}. In the standard picture the equilibrium tide is viewed as an oscillating large-scale shear flow in which smaller-scale convection damps it by acting as an effective viscosity $\nu_E$. An important issue is that the tidal forcing frequency ($\omega$) is often (much) larger than the turnover frequency of the dominant convective eddies ($\omega_c$) in applications \citep[e.g.][]{GO1997}. For example, the turnover timescale of convective eddies near the base of the solar convection zone is of order a month and the orbital period of short-period hot Jupiters is of order a day, leading to $\omega/\omega_c\sim 40\gg 1$ (the same regime applies also for eddies in the bulk). Phenomenological arguments by \citet{Zahn1966} and \citet{GN1977} suggested that $\nu_E$ should be reduced in the regime of fast tides when $\omega/\omega_c\gtrsim 1$ according to either $\nu_E\propto (\omega/\omega_c)^{-1}$ or $\nu_E\propto (\omega/\omega_c)^{-2}$.  It has recently become possible to test these scalings directly with numerical simulations \citep[starting with pioneering work by e.g.][]{Penev2009,OL2012}. Recent local \citep{OL2012,braviner_stellar_2015,DBJ2020,DBJ2020a} and global \citep{VB2020,VB2020a} numerical simulations have provided strong evidence that $\nu_E\propto (\omega/\omega_c)^{-2}$ for fast tides (though not for the reasons originally proposed), due to Reynolds stresses involving correlations between convective flow components. As a result, tidal evolution due to equilibrium tide dissipation is predicted to be weak in most applications involving pre- and main-sequence stars and giant planets \citep{DBJ2020a,B2020}, though it is probably still the dominant mechanism in giant stars \citep[e.g.][]{VerbuntPhinney1995,Mustill2012,Beck2018,SunArrasWeinberg2018,PriceWhelan2018}. 

Recently, \cite{Terquem2021} proposed an interesting way to think about the interaction between fast tides and convection. She employed a novel Reynolds decomposition of the flow into a mean and fluctuation using temporal averaging over an intermediate timescale (rather than spatial averaging, which is closer to the standard line of thought described above). The dominant terms in the (point-wise) interaction between tidal flows and convection were Reynolds stresses involving correlations of tidal flow components rather than convective flow components. This suggests a new term may dominate the interaction between fast tides and convection, referred to as $D_R$ (which will be described below). \cite{Terquem2021} made several assumptions to estimate $D_R$, indicating that it could enhance tidal dissipation of equilibrium tides to such an extent that it may explain the observed migration of the natural satellites of Jupiter and Saturn, and play a role in planetary orbital decay and binary circularization. 

Motivated by the importance of this problem for tidal dissipation, we revisit this issue and carefully determine the relevant terms in the interaction between tidal flows and convection. We demonstrate that the energy transfers associated with the new term proposed by \cite{Terquem2021} vanish identically in both Boussinesq and anelastic models of convection. This new result arises from considering \textit{spatial integration} and \textit{the correct (irrotational) equilibrium tide} -- rather than the equilibrium tide of \cite{Zahn1966}, which is usually invalid convection zones -- both of which were not properly treated in \cite{Terquem2021}. The new term proposed by \cite{Terquem2021} is therefore unfortunately unlikely to contribute significantly to dissipation of equilibrium tides.

The structure of this letter is as follows. In Section \S~\ref{EQMTIDE}, we introduce our model and review equilibrium tides in convection zones. We then go on to analyse each of the terms that contribute to the interaction between the tide and convection by deriving an energy equation for the convective flow in both Boussinesq (\S~\ref{Boussinesq}) and anelastic (\S~\ref{Anelastic}) models. We verify these results numerically with new hydrodynamical simulations in \S~\ref{Examples}, and conclude in \S~\ref{Conclusions}.

\section{Equilibrium tides in convection zones}
\label{EQMTIDE}

We consider a star (or planet) of mass $M$, radius $R$, and uniform rotation rate $\Omega$, with a convection zone that is either a spherical shell or a full sphere with volume $V$ and boundary $\partial V$. This configuration is relevant for the convection zones of MKGF stars and giant planet envelopes. We adopt spherical polar coordinates $(r,\theta,\phi)$ centred on the body, with the polar axis coinciding with the rotation axis. The body has a spherically-symmetric density distribution $\bar{\rho}(r)$ that is in hydrostatic equilibrium in the absence of tides, with inward gravitational acceleration $g(r)$. 

We assume efficient convection such that the body is barotropic. The conventional incompressible equilibrium tide of \cite{Zahn1966} does not then correctly describe the tidal response \citep{Terquem1998,GD1998} when $\omega^2\gtrsim -N_c^2$ (the squared buoyancy frequency), which is usually satisfied except near boundaries. Instead, the linear adiabatic equilibrium tidal flow ($\boldsymbol{u}_e$) is irrotational ($\nabla \times \boldsymbol{u}_e=\boldsymbol{0}$) in the fluid frame, defined by
\begin{eqnarray}
\label{EqmTide1}
\boldsymbol{u}_e=-\nabla \dot{X},
\end{eqnarray}
where $X$ is a potential and $\dot{X}$ is its time derivative. This satisfies
\begin{eqnarray}
\label{Xeqn}
\nabla \cdot(\bar{\rho} \nabla X) = \frac{1}{g}\frac{\mathrm{d}\bar{\rho}}{\mathrm{d}r}\Psi,
\end{eqnarray}
where $\Psi$ is the sum of the (quadrupolar) tidal potential and the perturbation to the gravitational potential of the body 
\citep{Ogilvie2013,B2020}. Boundary conditions match the irrotational equilibrium tide onto the conventional equilibrium tide in radiation zones or the (free) stellar surface i.e.~$\xi_{e,r}(r_b)=-\Psi(r_b)/g(r_b)$ (which does \textit{not} apply in the interior),
where $\partial_t\xi_{e,r}=u_{e,r}$ and $r_b$ is any boundary of a convection zone. Alternatively, in giant planets a solid core may instead be better modelled by $u_{e,r}(r_b)=0$ there.

The equilibrium tide has an associated linear Eulerian density perturbation that satisfies
\begin{eqnarray}
\label{rhoE}
\partial_t\rho'_e = -\nabla \cdot (\bar{\rho} \boldsymbol{u}_e),
\end{eqnarray}
and is neither incompressible ($\nabla \cdot \boldsymbol{u}_e\ne 0$) nor anelastic ($\nabla \cdot (\bar{\rho} \boldsymbol{u}_e)\ne 0$) for general $\bar{\rho}(r)$, such as in the anelastic convection problem we will discuss below. However, in a homogeneous body the equilibrium tide is incompressible.

To analyse the interaction between tidal flows and convection, we split the total flow ($\boldsymbol{u}$) into an equilibrium tide (subscript e) and the convection (subscript c) such that 
\begin{eqnarray}
\boldsymbol{u}=\boldsymbol{u}_e + \boldsymbol{u}_c,
\end{eqnarray}
together with $p=p_e+p_c$ for the pressure. Since tidal evolution proceeds much more slowly than convective timescales we treat $\boldsymbol{u}_e$ as perfectly maintained and probe the (instantaneous) energy transfers to determine the rates of tidal dissipation (which would modify the tide on much longer timescales). The primary interaction between the tide and convection comes about via 
the nonlinear advection term $\bar{\rho} \boldsymbol{u}\cdot \nabla\boldsymbol{u}$ in the momentum equation, which we will analyse below.

The equilibrium tide as defined here is strictly valid if $\omega^2\gg \Omega^2$. However, for any $\Omega$ we can define $\boldsymbol{u}_e$ as above, then Coriolis forces lead to wavelike/dynamical tides (plus possible non-wavelike corrections), which will also interact with convection but are not our focus here. These waves can be incorporated as part of $\boldsymbol{u}_c$, which would then consist of the convective plus wavelike tidal flows.

\section{Boussinesq convection}
\label{Boussinesq}

The simplest models of convection are Boussinesq/incompressible, in which density variations are accounted for only in the buoyancy term in the momentum equation \citep[e.g.][]{SV1960}. In such models, we consider incompressible fluid of uniform density $\bar{\rho}(r)=\rho$, with $\nabla \cdot \boldsymbol{u}_e=\nabla \cdot \boldsymbol{u}_c=0$. The momentum equation for the convection interacting with the equilibrium tide in the fluid frame is then
\begin{eqnarray}
\nonumber
&& \hspace{-0.3cm} \bar{\rho}\partial_t\boldsymbol{u}_c + \bar{\rho}\boldsymbol{u}_c\cdot \nabla \boldsymbol{u}_c + \bar{\rho}\boldsymbol{u}_e\cdot \nabla \boldsymbol{u}_c + \bar{\rho}\boldsymbol{u}_c\cdot \nabla \boldsymbol{u}_{e} + \bar{\rho}\boldsymbol{u}_e\cdot \nabla \boldsymbol{u}_e   \\ \label{BoussEq}
 && \hspace{0.3cm} +2\bar{\rho}\boldsymbol{\Omega}\times\boldsymbol{u}_c= -\nabla p_c +\bar{\rho}\boldsymbol{f}_e + \textrm{buoyancy}+ \textrm{viscous},
\end{eqnarray}
where $\boldsymbol{f}_e=-2\boldsymbol{\Omega}\times \boldsymbol{u}_e$ is the effective forcing of wavelike tides by the equilibrium tide, and $p_c$ includes the centrifugal potential. We construct an equation for the volume-integrated kinetic energy of the convection by taking the scalar product of Eq.~\ref{BoussEq} with $\boldsymbol{u}_c$ and integrating over the volume of the convection zone, with integration denoted by $\langle \cdot\rangle$. We obtain
\begin{eqnarray}
\label{Energy}
\partial_t(\langle \frac{1}{2}\bar{\rho}|\boldsymbol{u}_c|^2 \rangle) = \mathcal{I}_{cc} + \mathcal{I}_{ec} + \mathcal{I}_{ce} +\mathcal{I}_{ee} + \dots,
\end{eqnarray}
where the dots indicate work done by buoyancy and viscous forces not directly important for the present discussion, and the term 
$\langle \bar{\rho} \boldsymbol{u}_c \cdot \boldsymbol{f}_e\rangle$,
which describes energy transfers due to effective forcing of wavelike tides by equilibrium tides. The latter is not our focus here but will be important in rotating stars/planets and should be studied further. We focus on the nonlinear injection terms $\mathcal{I}_{..}$ which appear on the right-hand side, since some of these describe the interaction between equilibrium tides and convection. $\mathcal{I}_{cc}\equiv-\langle \bar{\rho}\boldsymbol{u}_c\cdot(\boldsymbol{u}_c\cdot \nabla \boldsymbol{u}_c)\rangle$ is not relevant for the interaction, but it can be shown to vanish using the divergence theorem, the boundary conditions and $\nabla \cdot \boldsymbol{u}_c=0$ (note also that $\langle \boldsymbol{u}_c\cdot \nabla p_c\rangle=0$ as well). We assume impenetrability $\boldsymbol{u}_c\cdot \boldsymbol{n}=0$ at convection zone boundaries in the bulge frame rotating with the orbit, where $\boldsymbol{n}$ is a normal vector to $\partial V$ (but $\boldsymbol{u}_e$ satisfies the conditions in \S~\ref{EQMTIDE}). This is appropriate at the surface when $\omega_c^2 \ll GM/R^3$ (valid except for the very fastest convective eddies in the lowest density surface layers), for a solid core, and for radiative/convective interfaces if the squared buoyancy frequency ($N^2$) in the stably-stratified layer satisfies $N^2\gg \omega_c^2$.

The tide-tide nonlinearity interaction with convection is described by
\begin{eqnarray}
\nonumber
\mathcal{I}_{ee}&\equiv&-\langle \bar{\rho}\boldsymbol{u}_c\cdot(\boldsymbol{u}_e\cdot \nabla \boldsymbol{u}_e)\rangle \\\nonumber &=& -\langle \bar{\rho}\boldsymbol{u}_c\cdot (\frac{1}{2}\nabla |\boldsymbol{u}_e|^2 -\underbrace{\boldsymbol{u}_e\times(\nabla\times \boldsymbol{u}_e)}_{=\boldsymbol{0}})\rangle
\\ &=& -\oint_{\partial V} \frac{1}{2}\bar{\rho}|\boldsymbol{u}_e|^2\boldsymbol{u}_c\cdot \mathrm{d}\boldsymbol{S},
\label{IeeEqn}
\end{eqnarray}
on application of the divergence theorem. This term vanishes after applying the boundary conditions on $\boldsymbol{u}_c$. To see this it is simplest to instead calculate this term in the bulge frame\footnote{Consider the ``equilibrium tidal flow" with Cartesian components $\boldsymbol{u}_e=\gamma(-ay/b, bx/a,0)$ in the bulge frame rotating with the orbital frequency, for a circularly orbiting, aligned, asynchronously rotating homogeneous star with ellipsoidal surface $x^2/a^2+y^2/b^2+z^2/c^2=R^2$, where $\gamma$ is the difference between the spin and orbital frequencies and $a, b$ and $c$ are the semi-axes \citep{BBO2016}. This is not irrotational and has a uniform vorticity component, but it is an exact solution satisfying $\mathcal{I}_{ee}=0$ because $\boldsymbol{u}_e\cdot\nabla \boldsymbol{u}_e=-\frac{\gamma^2}{2}\nabla (x^2+y^2)$, together with $\boldsymbol{u}_e\cdot\boldsymbol{n}=0$, where $\boldsymbol{n}=(x/a^2,y/b^2,z/c^2)$. Specifically, we have $\boldsymbol{u}_c\cdot\boldsymbol{n}=0$ in the bulge frame to leading order if $\omega_c^2\ll GM/R^3$, implying $\mathcal{I}_{ee}=\oint_{\partial V} \bar{\rho}\frac{\gamma^2}{2} (x^2+y^2) \boldsymbol{u}_c\cdot\boldsymbol{n}\,\mathrm{d}S=0$ in this frame.}. It also vanishes in the fluid frame on time-averaging over a tidal period for fast (linear) tides, in which $\boldsymbol{u}_c$ is approximately steady, since the boundary is periodically-deformed.
Hence there is no energy exchange between the tidal and convective flows as a result of this term. We can also write
\begin{eqnarray}
\nonumber
\mathcal{I}_{ee}&=&-\langle \bar{\rho}\nabla \cdot ((\boldsymbol{u}_c\cdot \boldsymbol{u}_e)\boldsymbol{u}_e )\rangle+\langle \bar{\rho}\boldsymbol{u}_e\cdot (\boldsymbol{u}_e\cdot \nabla) \boldsymbol{u}_c\rangle, \\
&=&\hspace{-0.25cm}-\underbrace{\oint_{\partial V} (\bar{\rho}(\boldsymbol{u}_c\cdot \boldsymbol{u}_e)\boldsymbol{u}_e )\cdot\mathrm{d}\boldsymbol{S}}_{F}+\underbrace{\langle \bar{\rho}\boldsymbol{u}_e\cdot (\boldsymbol{u}_e\cdot \nabla) \boldsymbol{u}_c\rangle}_{=\langle \bar{\rho} D_R\rangle},
\label{Fterm}
\end{eqnarray}
which relates $\mathcal{I}_{ee}$ to $D_R$ of \cite{Terquem2021}. However, they defined $D_R$ before spatial integration, and also performed a time average. This last term involving tide-tide correlations and gradients of the convective flow was proposed to be dominant for fast tides by \cite{Terquem2021}.

The flux term $F$ is identically zero in the bulge frame in which $\boldsymbol{u}_e\cdot\boldsymbol{n}=0$ (see footnote 1),
and in giant planets with cores satisfying no-slip conditions on which $\boldsymbol{u}_c=\boldsymbol{0}$ (as also in idealized spherical simulations like those in \citealt{VB2020,VB2020a}, but it may not then vanish with stress-free conditions).
Since $\mathcal{I}_{ee}$ vanishes, this implies $\langle \rho D_R\rangle =0$ also. Hence, the new term proposed by \cite{Terquem2021} vanishes identically after spatial integration in Boussinesq convection and cannot contribute to tidal dissipation in such models. Another perspective is that $\boldsymbol{u}_e\cdot \nabla \boldsymbol{u}_e=(1/2)\nabla |\boldsymbol{u}_e|^2$ 
only modifies the pressure, which has no effect beyond simply deforming the boundary in incompressible models.

We also find 
\begin{eqnarray}
\nonumber
\mathcal{I}_{ec}&\equiv&-\langle \bar{\rho}\boldsymbol{u}_c\cdot(\boldsymbol{u}_e\cdot \nabla \boldsymbol{u}_c)\rangle = -\langle \bar{\rho}\nabla\cdot (\frac{1}{2}|\boldsymbol{u}_c|^2\boldsymbol{u}_e) \rangle \\ &=& -\oint_{\partial V} (\frac{1}{2}\bar{\rho}|\boldsymbol{u}_c|^2\boldsymbol{u}_e)\cdot \mathrm{d}\boldsymbol{S}=0,
\end{eqnarray}
since $\nabla \cdot \boldsymbol{u}_e=0$ in incompressible/Boussinesq models, which vanishes for similar reasons to $F$ above.

As a result, \textit{the only term contributing to the interaction between equilibrium tides and Boussinesq convection} is
\begin{eqnarray}
\mathcal{I}_{ce} \equiv -\langle \bar{\rho}\boldsymbol{u}_c\cdot(\boldsymbol{u}_c\cdot \nabla \boldsymbol{u}_e)\rangle,
\end{eqnarray}
i.e.~ Reynolds stresses involving correlations between convective flow components. $\mathcal{I}_{ce}$ fully characterises the interaction between equilibrium tides and convection in incompressible/Boussinesq models.
This term was analysed in detail with numerical simulations in local models by \cite{OL2012,DBJ2020,DBJ2020a} and idealized spherical Boussinesq models by \cite{VB2020,VB2020a}. While $\mathcal{I}_{ce}$ vanishes for asymptotically fast tides ($|\omega|/\omega_c\to\infty$) under Terquem's Reynolds decomposition, which is consistent with an effective viscosity $\nu_E\propto \omega^{-2}$, it is in general small \textit{but nonzero} for realistic finite values of $|\omega|/\omega_c\gg 1$.

Note that \cite{Terquem2021} referred to the integrand of $\mathcal{I}_{ce}$ as $D_R^{st}$ (before spatial integration), and attempted to show that this term was much smaller than $D_R$ in stellar and planetary models. While the point-wise magnitude of $\rho D_R$ could potentially be larger for fast tides, we have demonstrated that $\langle \rho D_R\rangle$ vanishes identically in incompressible/Boussinesq models after integration.
\cite{Terquem2021} did not perform spatial integration until after approximating this term with a typical magnitude, first assuming it to be positive everywhere. We have demonstrated analytically that this is incorrect due to its omission of cancellations. They also adopted the (incorrect) conventional equilibrium tide of \cite{Zahn1966}, which is not irrotational. These reasons may explain why the above result was previously missed.

Since stars and planets are not truly incompressible (though incompressible models were studied by all of the above-mentioned works) it is important to ask: how does this result carry over to more realistic models? To make progress towards answering this we turn to consider the interaction between tidal flows and anelastic convection.

\section{Anelastic convection}
\label{Anelastic}

In the anelastic approximation we continue to assume slow flows relative to the sound speed, but allow variations in density in the domain such that $\bar{\rho}(r)$ in the unperturbed star/planet. Anelastic models are widely used to study stellar and planetary convection \citep[e.g.][]{Jones2011}. The momentum equation for the convection is Eq.~\ref{BoussEq} except that $\bar{\rho}(r)$ is not constant and we also have $\nabla \cdot (\bar{\rho} \boldsymbol{u}_c)=0$. On taking the scalar product of Eq.~\ref{BoussEq} with $\boldsymbol{u}_c$ and performing spatial integration, the energy equation for the convective flow looks similar to Eq.~\ref{Energy} except that $\bar{\rho}$ is not constant. Once again $\mathcal{I}_{cc}=0$, on application of the anelastic constraint here. Energy transfers by the tide-tide nonlinearity satisfy
\begin{eqnarray}
\mathcal{I}_{ee}&=&
-\oint_{\partial V} \frac{1}{2}\bar{\rho}|\boldsymbol{u}_e|^2\boldsymbol{u}_c\cdot \mathrm{d}\boldsymbol{S},
\end{eqnarray}
which vanishes for the same reasons\footnote{Since $\boldsymbol{u}_e=-\nabla \dot{X}(r,\theta,\phi,t)$ in the fluid frame, there is an associated steady flow in the bulge frame rotating at $n\boldsymbol{e}_z$ (with quantities denoted by primes) for an aligned circular orbit $\boldsymbol{u}_e'=-\nabla' \dot{X}(r',\theta',\phi') + \gamma\boldsymbol{e}_z\times\boldsymbol{x}'$, and $\gamma=\Omega-n$. It can be shown that $\boldsymbol{u}_e'\cdot \nabla' \boldsymbol{u}_e'=\nabla' \chi - \gamma\boldsymbol{e}_z\times\nabla' \dot{X}$ for a scalar field $\chi$. The latter term contributes only to the Coriolis acceleration defining $\boldsymbol{u}_e'$ and the former contributes as a surface integral in $\mathcal{I}_{ee}$ (and hence vanishes using $\boldsymbol{u}_c'\cdot \boldsymbol{n}'=0$), and we also have $\boldsymbol{u}_e'\cdot\boldsymbol{n}'=0$ (normal vector $\boldsymbol{n}'$) on the surface for general $\bar{\rho}(r)$.} 
as Eq.~\ref{IeeEqn} except that the convection satisfies $\nabla \cdot (\bar{\rho}\boldsymbol{u}_c)=0$. In reality $\bar{\rho}\to 0$ at the surface also, further justifying that this integral vanishes. Hence, there is no energy exchange between the convective and tidal flows permitted by this term.
We can also rewrite this term as
\begin{eqnarray}
\nonumber
\mathcal{I}_{ee}&=&-\underbrace{\langle \nabla \cdot ((\boldsymbol{u}_c\cdot \boldsymbol{u}_e)\bar{\rho}\boldsymbol{u}_e)\rangle}_{=0}+ \langle (\boldsymbol{u}_c\cdot \boldsymbol{u}_e)\nabla \cdot (\bar{\rho}\boldsymbol{u}_e)\rangle\\ && \hspace{1cm} +
\underbrace{\langle\bar{\rho}\boldsymbol{u}_e\cdot (\boldsymbol{u}_e \cdot \nabla \boldsymbol{u}_c)\rangle}_{=\langle \bar{\rho} D_R\rangle},
\end{eqnarray}
where the flux term vanishes as in Eq.~\ref{Fterm}. Since $\mathcal{I}_{ee}=0$, we have the exact balance (using Eq.~\ref{rhoE})
\begin{eqnarray}
\label{Balance}
\langle\bar{\rho}\boldsymbol{u}_e\cdot (\boldsymbol{u}_e \cdot \nabla \boldsymbol{u}_c)\rangle=\langle (\boldsymbol{u}_c\cdot \boldsymbol{u}_e)\partial_t\rho_e'\rangle.
\end{eqnarray}
This indicates that even though $\langle \bar{\rho} D_R\rangle$ is nonzero in the anelastic case, it is exactly balanced and does not exchange energy with the convective flow because $\mathcal{I}_{ee}$ vanishes.

We also find
$\mathcal{I}_{ec}
= -\langle \frac{1}{2}|\boldsymbol{u}_c|^2 \partial_t \rho'_e \rangle\ne 0
$
in general since the equilibrium tide is not anelastic (the flux term vanishes as in Eq.~\ref{Fterm}). This describes energy transfers due to advection of convective flow energy per unit mass by compressional motion of equilibrium tides, which balances a term in the rate of change of total kinetic energy $(1/2)\langle(\bar{\rho}+\rho_e')|\boldsymbol{u}|^2\rangle$, and we note that $\langle\bar{\rho}\boldsymbol{u}_c\cdot\boldsymbol{u}_e\rangle=0$ using Eq.~\ref{EqmTide1}).

As a result, the interaction between equilibrium tides and convection is likely determined by
\begin{eqnarray}
\mathcal{I}_{ce} = -\langle \bar{\rho}\boldsymbol{u}_c\cdot(\boldsymbol{u}_c\cdot \nabla \boldsymbol{u}_e)\rangle,
\end{eqnarray}
just like in the Boussinesq case in \S~\ref{Boussinesq} (with a possible contribution from $\mathcal{I}_{ec}$ and terms proportional to $\rho_e'$). $\mathcal{I}_{ce}$ is the anelastic generalization of $D_R^{st}$ (after multiplying by $\bar{\rho}$ and performing spatial integration) in \cite{Terquem2021} and this term should be studied in future simulations to build upon its study in the Boussinesq case by e.g.~\cite{OL2012,DBJ2020,DBJ2020a,VB2020,VB2020a}.

We have thus confirmed that the term proposed by \cite{Terquem2021} is also unlikely to contribute to dissipation of equilibrium tides in more realistic anelastic models.

\section{Illustrative hydrodynamical simulations}
\label{Examples}

To verify our results, we present new proof-of-concept hydrodynamical simulations of both Boussinesq and (Lantz-Braginsky-Roberts) anelastic non-rotating convection interacting with (quadrupolar) equilibrium tides in a spherical shell. We have modified the spherical pseudo-spectral code MagIC 5.10 \citep{Wicht2002,GastineWicht2012} to solve Eq.~\ref{BoussEq}, along with Eq.~\ref{Xeqn} for $X$ assuming a rigid core. These constitute the first global anelastic simulations and the first in shells for this problem. We adopt spherical boundaries at $r=r_b=0.5$ and $r=R=1$ on which $\boldsymbol{u}_c=\boldsymbol{0}$ to ensure the flux terms in \S~\ref{Boussinesq} and \ref{Anelastic} vanish, with fixed entropy $S=0$ at the top and fixed flux $\partial_r S=-1$ at the bottom. We use constant kinematic viscosity $\nu$ equal to the thermal diffusivity $\kappa$, and specific heat $c_p$, with viscous time units, and unity outer boundary density. The anelastic case is a centrally-condensed polytrope with $N_\rho=3$ density scale heights (factor of 20 variation in density) and polytropic index $m=3/2$ \citep[e.g.][]{Jones2011} in which $X$ is computed numerically, whereas the Boussinesq case has $N_\rho=0$, both with $g=g_0(R/r)^{2}$. The flux based Rayleigh number is $\mathrm{Ra}=g_0 (-\partial_r S) R^4/(\nu\kappa c_p)$.

We crudely define $\omega_c=\sqrt{\langle u_{c,r}^2\rangle}/(R-r_b)$ here for $N_\rho=0$ and $\omega_c=\sqrt{\langle u_{c,r}^2\rangle}/((R-r_b)/3)$ for $N_\rho=3$ (even if it depends on $r$ in the latter). 
Both simulations have $\mathrm{Ra}=10^6$, a dimensionless tidal amplitude $A=0.05$ for a circular aligned orbit \citep[e.g.][]{VB2020a} with frequency $\omega=1000$ for $N_\rho=0$ or $\omega=5000$ for $N_\rho=3$. We find $\omega_c\approx 62.1$ when $N_\rho=0$ or $\omega_c\approx 478$ when $N_\rho=3$, so that $\omega/\omega_c\approx 16.1$ or $\omega/\omega_c\approx 10.5$, i.e.~we consider fast tides.

\begin{figure*}
  \begin{center}
    \subfigure{\includegraphics[
    trim=0cm 0.4cm 1.93cm 0.98cm,clip=true,
    width=0.33\textwidth]{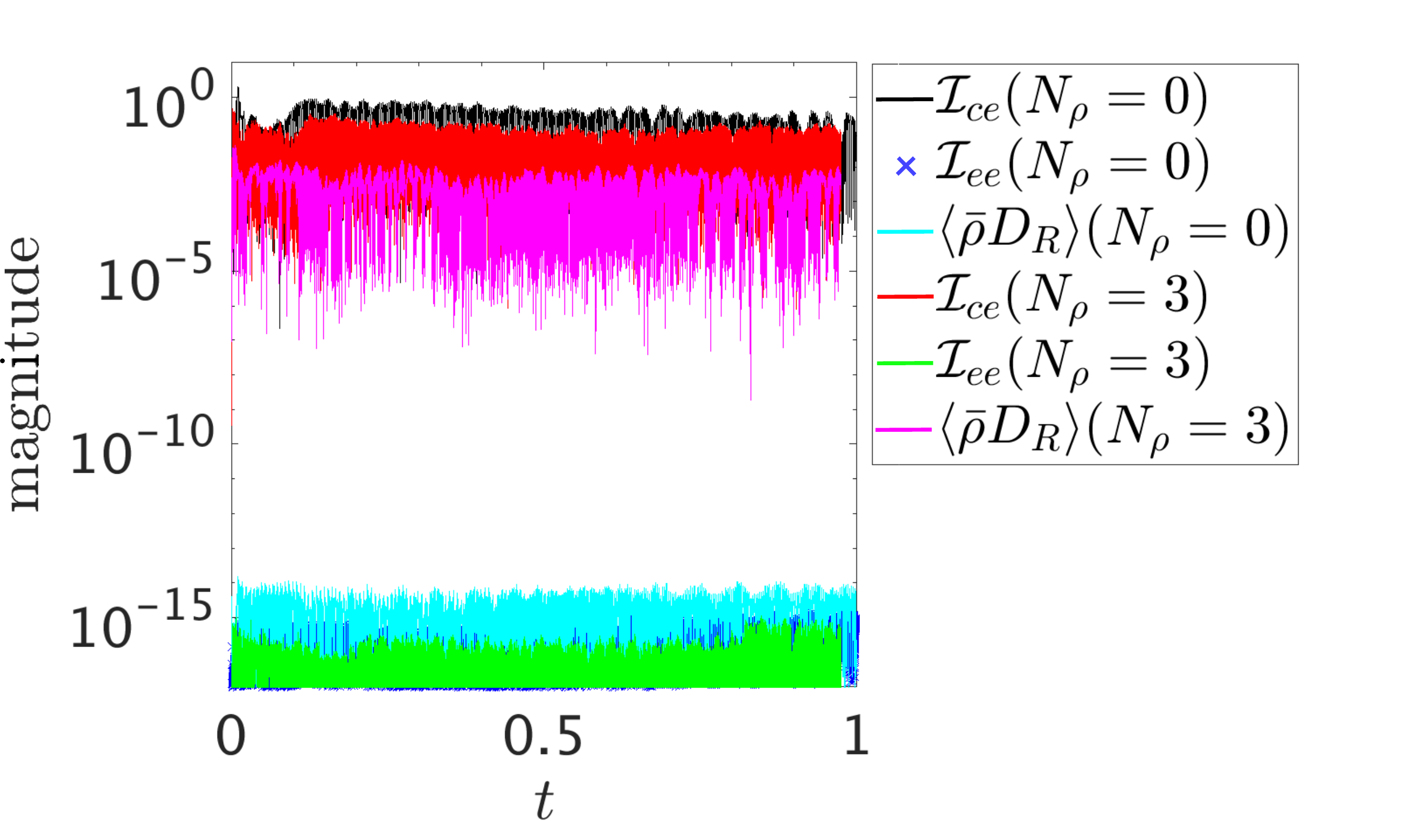}}
    \subfigure{\includegraphics[
    trim=0cm 0.4cm 2.3cm 0.2cm,clip=true,
    width=0.33\textwidth]{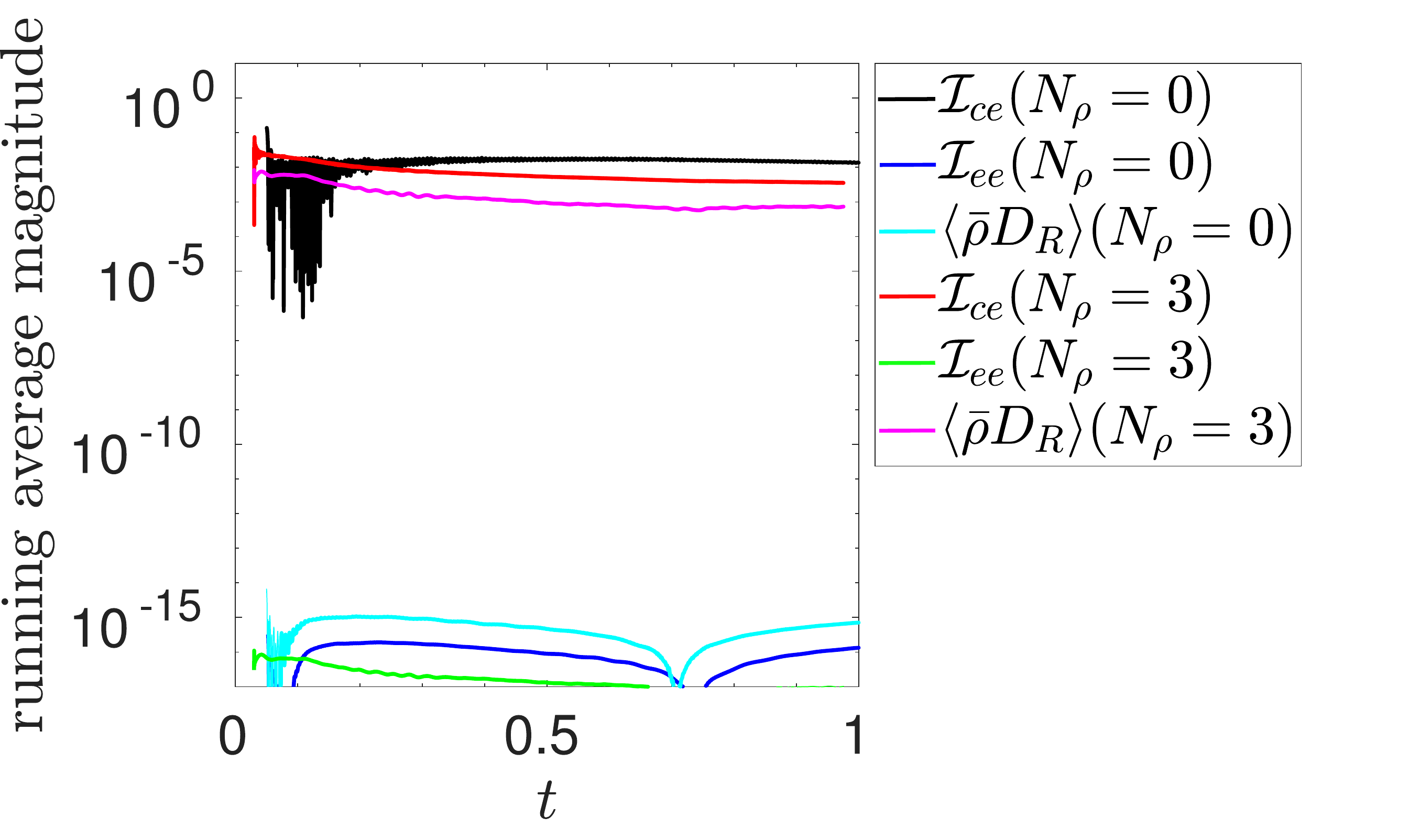}}
    \subfigure{\includegraphics[
    trim=0cm 0.4cm 2.cm 1cm,clip=true,
    width=0.33\textwidth]{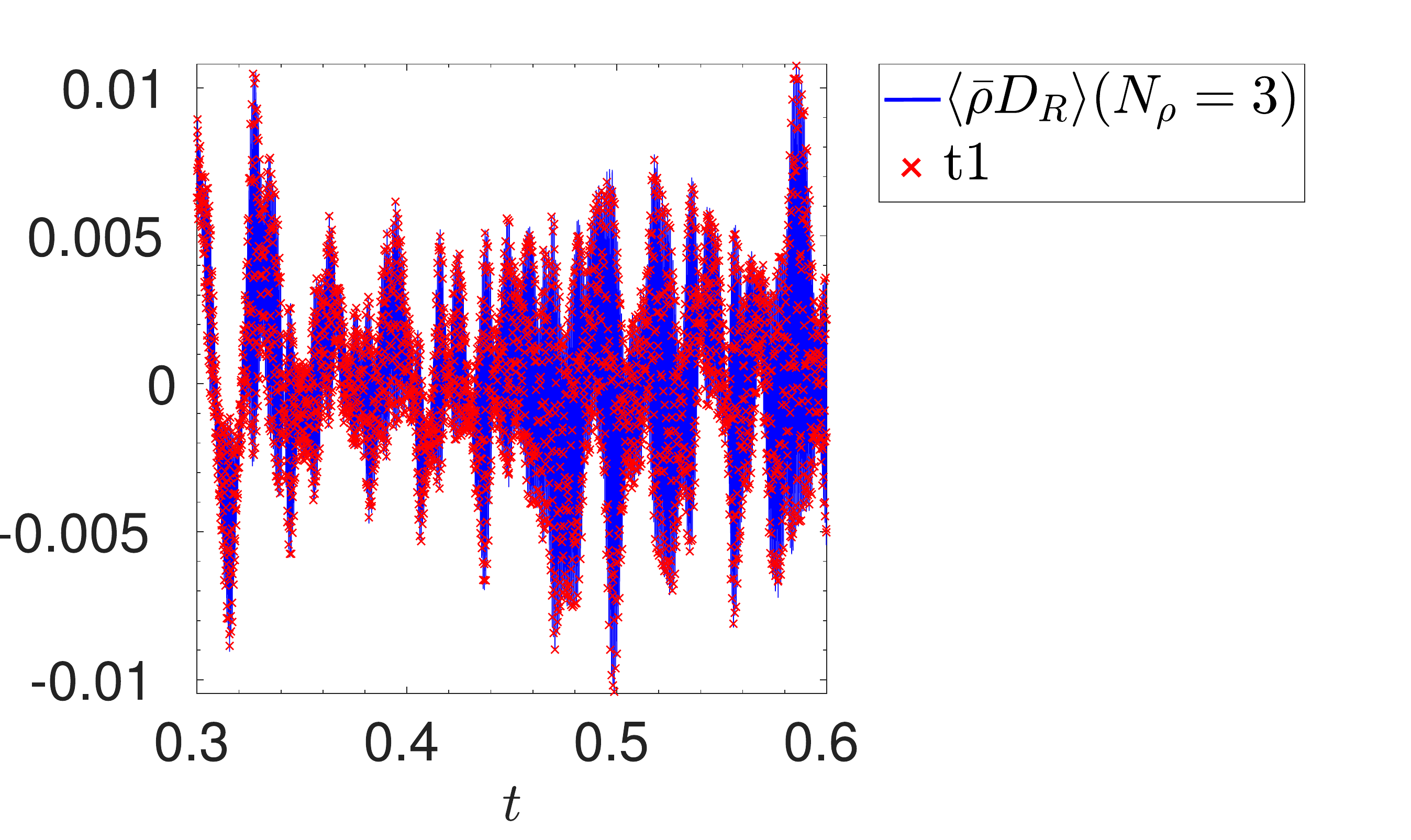}}
     \end{center}
  \caption{Numerical evaluation of $|\mathcal{I}_{ce}|$ (normalized by $\bar{\rho}(r_b) u_c^2 A\omega$, where $u_c$ is the root-mean-square radial velocity), $|\mathcal{I}_{ee}|$ and $|\langle \bar{\rho}D_R\rangle|$ (both normalized by $\bar{\rho}(r_b) u_c A^2\omega^2R$) in a Boussinesq $(N_\rho=0)$ and an anelastic $(N_\rho=3)$ model. Left: quantities as a function of time (viscous units). Middle: running time averages. Right: anelastic case showing the exact balance  (normalized by $\bar{\rho}(r_b) u_c A^2\omega^2R$) in Eq.~\ref{Balance}, where t1 is the term on the right hand side. Spatial resolution: $97$ Chebyshev points in radius and spherical harmonics up to degree $85$.}
  \label{Fig1}
\end{figure*}

We show in Fig.~\ref{Fig1} results from these illustrative simulations. The left (and middle) panels verify that $\mathcal{I}_{ce}$ (and its running average) rather than $\mathcal{I}_{ee}$ is responsible for the interaction between tidal flows and convection, that $\langle \bar{\rho} D_R\rangle$ is consistent with zero in the Boussinesq case and (right panel) satisfies the exact balance given by Eq.~\ref{Balance} in the anelastic case. In the latter, the time average of $|\langle \bar{\rho}D_R\rangle|$ is nonzero but is much smaller than the value obtained by estimating it point-wise by a typical magnitude before spatial integration (by a factor of more than $10^{3}$). In both simulations $\mathcal{I}_{ee}$ (and its running average) is consistent with zero, with a tiny nonzero value due to numerical errors. Further details of these simulations will be presented elsewhere.

\section{Conclusions}
\label{Conclusions}

We have revisited the interaction between equilibrium tides and convection by carefully 
considering the relevant terms contributing to this interaction in two different convective models. 
This work was motivated by the tantalizing suggestion by \cite{Terquem2021} that a previously-neglected term could dominate this interaction based on applying a novel Reynolds decomposition of the flow. In this paper we have studied this term using both analytical arguments and numerical simulations of both Boussinesq and anelastic convection in spherical geometry. We demonstrated analytically that this term vanishes identically when the (correct) irrotational equilibrium tide is employed in convection zones and when spatial integration is performed. Our numerical simulations have confirmed this result. Hence, the new term proposed by \cite{Terquem2021} is unlikely to contribute to equilibrium tide dissipation in stars and planets.

Our arguments generally apply for fast tides, but do not depend on sizes of convective eddies, or on the anisotropy of convection. They also do not depend on (differential) rotation (including centrifugal effects) as long as we define the equilibrium tide as in \S~\ref{EQMTIDE}. However, tidal forcing can excite inertial waves/modes in convection zones with rotation \citep[e.g.][]{PapIv2010,Ogilvie2013,FBBO2014,Mathis2015,B2020}, which exchange energy with convection through $\langle\bar{\rho} \boldsymbol{u}_c\cdot\boldsymbol{f}_e\rangle$, which should be studied in future work. We have ignored non-adiabatic effects in modifying equilibrium tides in near surface layers \citep[e.g.][]{Bunting2019}, but this is unlikely to change our conclusions on account of the low stellar density there, where the fast tide regime also does not apply in any case. Anelastic models are also invalid in near surface layers, but they are believed to adequately describe the bulk of convection zones. 

While Reynolds stresses involving tide-tide correlations are likely unimportant for equilibrium tide dissipation owing to the high degree of cancellation involved when spatially integrating these contributions, the formalism of \cite{Terquem2021} could potentially be fruitfully applied to study interactions between waves and convection e.g.~inertial waves or stochastically excited p-modes.

The interaction between equilibrium tides and convection is therefore likely to be dominated by terms considered in previous Boussinesq models \citep[e.g.][]{OL2012,DBJ2020a,VB2020a}. These usually indicate weak dissipation of equilibrium tides, such that tidal dissipation in pre- and main-sequence stars and giant planets is probably instead due to inertial and internal gravity waves \citep[e.g.][]{B2020} -- though equilibrium tides are likely to be dominant in giant stars. A major uncertainty remains in the application of Boussinesq results to stellar models. Ultimately we require simulations using compressible or anelastic formulations that capture the dynamics of several scale heights. We have begun this line of investigation here (building upon early work by \citealt{Penev2009}), but much work is left to quantify and understand the interaction between tidal flows and convection.
\vspace{-0.5cm}
\section*{Acknowledgements}
\vspace{-0.2cm}
Research supported by STFC grants ST/R00059X/1 and 
ST/S000275/1. Simulations undertaken on ARC4, part of the High Performance Computing facilities at the University of Leeds. We thank Thomas Gastine for help with MagIC and Craig Duguid, Jeremy Goodman, Pavel Ivanov, Gordon Ogilvie, John Papaloizou (the referee) and Caroline Terquem for comments that helped clarify our presentation.
\vspace{-1.2cm}
\section*{Data availability}
\vspace{-0.2cm}
The data underlying this article will be shared on reasonable request to the corresponding author. MagIC 5.10 website: \url{https://magic-sph.github.io/}.
\vspace{-0.8cm}
\bibliography{tid}
\bibliographystyle{mnras}
\label{lastpage}
\end{document}